\documentclass[12pt]{article}
\usepackage[dvips]{graphicx}
\begin{document}
\newcommand{\Rs}{R_{\odot}}
\title{Reply to the Comments of Dikpati et al.}
\author{Arnab Rai Choudhuri$^{1}$, Dibyendu Nandy$^{2}$, \and 
Piyali Chatterjee$^{1}$}
\maketitle

{$^{1}$Department of Physics, Indian Institute of Science, 
Bangalore-560012, India. \\
$^{2}$Department of Physics, Montana State University, Bozeman, MT 59717, 
USA.}\\
\section*{Abstract}
We present here our response to Dikpati et al.'s criticism of our
recent solar dynamo model.

\section{Introduction}
Dikpati et al.\ (2005; hereafter DRGM) have written a comment on a
recent paper by us (Chatterjee, Nandy and Choudhuri 2004; hereafter
CNC) presenting a solar dynamo model.  The criticisms of DRGM 
broadly fall under two categories.
Firstly, they point out that they are unable to reproduce our results
(see \S3.2, \S3.3 and \S3.5 of their paper).  Secondly, they have
raised some concerns about the basics of our model, including the
name `circulation-dominated dynamo' given by us to our model (see \S3.1, 
\S3.4, \S4.1 and \S4.2 of DRGM).  These two different kinds of criticisms
are addressed in \S2 and \S3 of this paper respectively.

\section{Possible reason for divergent results}

Unfortunately there is a typographical error in eq.\ (11) of CNC,
which gives the stream function used to generate the velocity field.
The correct expression of the stream function which is implemented
in our code is
\begin{eqnarray}
\psi r \sin \theta = \psi_0 (r - R_p) \sin \left[ \frac{\pi (r - R_p)}
{(\Rs - R_p)} \right] \{ 1 - e^{- \beta_1 \theta^{\epsilon}}\} \\ \nonumber
\{1 - e^{\beta_2 (\theta - \pi/2)} \} e^{-((r -r_0)/\Gamma)^2}
\end{eqnarray}
with the following values of the parameters: 
$\beta_1 = 1.36$, $\beta_2 = 1.63$, 
$\epsilon = 2.0000001$, $r_0 = (\Rs - R_b)/4.0$, $\Gamma = 
3.47 \times 10^{8}$ m, $\gamma = 0.95$, $m=3/2$. It was mistakenly
printed as
\begin{eqnarray}
\psi r \sin \theta = \psi_0 (r - R_p) \sin \left[ \frac{\pi (r - R_p)}
{(\Rs - R_p)} \right] \{ 1 - e^{- \beta_1 r \theta^{\epsilon}}\} \\ \nonumber
\{1 - e^{\beta_2 r (\theta - \pi/2)} \} e^{-((r -r_0)/\Gamma)^2}
\end{eqnarray}
with
$\beta_1 = 1.36 \times 10^{-8}$ m$^{-1}$ and $\beta_2 = 1.63 \times 10^{-8}$ 
m$^{-1}$, while the values of the other parameters were given correctly as
quoted above.  Whereas the parameters $\beta_1$ and $\beta_2$ in (1) are
dimensionless quantities (as they are in our code), they have dimensions
of inverse length in (2).  We had used stream functions of the form (2),
though not exactly the stream function (2),
in some of our earlier works (Dikpati \& Choudhuri 1995; Choudhuri,
Sch\"ussler \& Dikpati 1995; Choudhuri \& Dikpati 1999; Nandy \&
Choudhuri 2001). Subsequently, however, we found that a stream function
of the form (1) gives more satisfactory results in kinematic dynamo
models.  The stream function (1) is used in the papers by Nandy \&
Choudhuri (2002), CNC (2004) and Choudhuri, Chatterjee \& Nandy (2004).  
While preparing the texts, we were cutting and pasting various things 
from the LaTex files of earlier papers and the wrong expression for the
stream function inadvertently crept in both in the {\em Materials and
Methods} of Nandy \& Choudhuri (2002) and in the CNC paper.  
We sincerely regret this and are extremely grateful 
to DRGM for their efforts
in reproducing our results, which made us aware of this typographical
mistake. However, we do not think that DRGM's use of a slightly 
different meridional circulation caused the difference between our results.
Our code gives qualitatively similar results with both the stream
functions (1) and (2). Another typographical mistake in CNC is that
the value of $r_{\rm TCZ}$ appearing in (13) is given as $0.95\Rs$
rather than $0.975\Rs$ which is the value used.

We believe that DRGM's result differs from ours because of their
unsatisfactory handling of magnetic buoyancy.  We have made some runs
by switching off our buoyancy algorithm and using the non-local buoyancy
of Dikpati \& Charbonneau (1999), i.e.\ multiplying $\alpha$ not by
the local $B$ but by $B(r = 0.7 R_{\odot}, \theta)$. In contrast
to the results presented in CNC, we get 
multi-lobed patterns, as seen in the snapshot of magnetic fields
shown in Fig.~1. This figure is very similar to Fig.~3 of DRGM.
The period for this dynamo solution comes
out to be 6.1 yr. In the buoyancy algorithm of CNC, only when $B$ is larger
than a critical value $B_c$ above $r=0.71\Rs$, the toroidal field
erupts to the surface and contributes to the generation of the 
poloidal field.  This happens only at sufficiently low latitudes.
In contrast, the non-local buoyancy algorithm of
Dikpati \& Charbonneau (1999) makes even a weak toroidal
field $B(r = 0.7 R_{\odot}, \theta)$ at high latitudes
contribute to the generation of poloidal field at the surface---a
clearly unphysical mechanism which upsets the dynamo solution
completely. The period of the dynamo also becomes shorter 
because the toroidal field starts generating poloidal field
while still at mid-latitudes, instead of having to be advected all the
way to low latitudes.

\begin{figure}
\centering{\includegraphics[width=7.0cm,height=6cm]{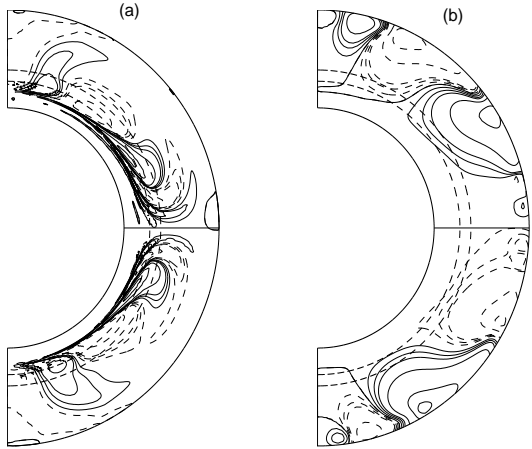}}
\caption{Toroidal (left) and poloidal (right) magnetic fields
in the dynamo solution obtained with the non-local buoyancy
formalism of Dikpati \& Charbonneau (1999).}   
\end{figure}

We plan to make our dynamo code available in the public domain, with
the settings of parameters used to generate our standard solution
presented in \S3.1 and \S4 of CNC.  We are now in the process of
making the code more user-friendly and preparing a guide for it.
The code and the guide should be available in the public domain
latest by 1 December 2005.  The entire solar physics community should be able
to examine our code and verify whether it produces the results presented
in CNC.  

\section{Response to other criticisms}

We now respond to the other criticisms of DRGM on the basics of our 
model---criticisms which do not necessarily depend on the fact that DRGM were 
unable to reproduce our results.

\subsection{Is our dynamo model circulation-dominated?}  

The longest subsection of DRGM (\S3.1) is devoted to questioning the
name `circulation-dominated' we had given to our dynamo model. Although
the discussion of DRGM seems to us nothing more than mere quibbling
over semantics,  let us explain our point of view.  We have $d \Omega/
dr$ positive at the lower latitudes, and the $\alpha$ coefficient
is also positive.  According to the well-known dynamo sign rule (see,
for example, Choudhuri 1998, \S16.5), the dynamo wave should propagate
poleward.  Still the toroidal field below the bottom of the solar convection
zone (SCZ) moves equatorward, because it is advected through a region
where the diffusivity has a low value $\eta_{RZ} = 2.2 \times 10^8$
cm$^2$ s$^{-1}$ and it essentially remains frozen in the fluid.  Even
if we take a rather low value of $d = 10^4$ km for the thickness of the
layer below the SCZ through which the toroidal field is advected, still
the diffusion time $d^2/\eta_{RZ}$ turns out to be about 144 yr---much
larger than the time scale of advection by the meridional circulation.
That is why the toroidal field simply gets advected through a thin
layer below the bottom of SCZ while remaining frozen and we have felt
that `circulation-dominated' is the appropriate name of a dynamo model
in which this happens.  If DRGM do not like our name, they are free to
use any other name.  We are surprised that they took about one journal
page to debate something which appears to us a trivial matter of semantics.

\subsection{TF/PF ratios}

In \S3.4 of DRGM, our model is criticized on the ground that polar fields
of order 2 kG are needed to generate 100 kG toroidal fields.  The answer
to this criticism can be found in \S2--3 of an already published paper by
Choudhuri (2003).  DRGM's argument is based on a serious misconception.
The dynamo equation deals with the mean magnetic field.  On the other hand,
flux tube rise simulations suggested a magnetic field of 100 kG only inside
the flux tubes---the mean field being much less if the flux is organized
intermittently. Choudhuri (2003, \S2) presented some straightforward 
back-of-the-envelope estimates showing that a circulation-dominated
dynamo can stretch a polar field of 10 G to a maximum toroidal field
of only about 10 kG.  So the mean toroidal field can be at most of this
value.  The toroidal field has to be highly intermittent, with the value
of 100 kG occuring only in isolated regions.  This is an entirely
consistent physical scenario.

\subsection{Different diffusivities for PF and TF}

Since there are not separate conducting fluids for toroidal and poloidal
fields, DRGM argue in \S4.2 that assuming different diffusivities at the
same point is implausible.  However, the diffusivities entering the dynamo
equation are not actual `physical' diffusivities, but effective turbulent 
diffusivities which arise from an averaging procedure 
and describe how the mean fields
evolve.  We believe that the magnetic field at the base of SCZ looks as
sketched in Fig.~3 of Choudhuri (2003), with the flux concentrations stretching
along the $\phi$ direction. Turbulent diffusivity is obviously suppressed
within these flux concentrations, making the magnetic field $B_{\phi}$
in these regions of concentration rather immune to turbulent diffusivity.
If we were doing calculations with the full magnetic field, then we could
capture this effect through a simple quenching of turbulent diffusivity.
But we have to average over regions much larger than the flux concentrations
to get the mean field dynamo equation.  How do we capture the information
in the dynamo equation that diffusivity is suppressed within $B_{\phi}$
concentrations having sizes smaller than the scale of our mean field theory?
We felt that taking a smaller diffusivity for $B_{\phi}$ is one way of
handling this issue.  We do not claim that this is a very satisfactory or
mathematically rigorous way.  But we have not been able to think up
any better way of tackling this issue.

\subsection{Penetration depth of meridional circulation}

DRGM argue in \S4.1 that the meridional circulation could not penetrate
below the bottom of SCZ which we require. It may be pointed out that
Dikpati \& Charbonneau (1999) produced their best dynamo models with
a deeply penetrating meridional circulation, without
clearly mentioning anywhere in the text
that this was essential to obtain good models.
The fault of Nandy \& Choudhuri (2002) seems to be that they were the
first who asserted clearly that a penetrating flow is required to produce
satisfactory circulation-dominated solar dynamo models and discussed
its physical implications. Fig.~2 shows the radial profiles of $v_{\theta}$
at $\theta= 45^{\circ}$ used by us as well as by Dikpati \& Charbonneau (1999).
No comment is necessary.  Linguists and sociologists of science would
be particlularly intrigued to note the language 
employed by DRGM to describe these
two velocity fields.  DRGM write ``Dikpati \& Charbonneau (1999) used
an equatorward return flow which penetrated slightly below $0.7R$'',
while DRGM's comment on our velocity field is: ``Nandy \& Choudhuri (2002)  
had a subsurface return flow of about a 10 m s$^{-1}$ at $r=0.6R$''!
We have no clue whatsoever how DRGM arrived at the
extraordinary conclusion of our model having a velocity of
10 m s$^{-1}$ at $r = 0.6R$, when a plot of velocity versus depth
was given in Fig.~3 of CNC.
\begin{figure}
\centering{\includegraphics[width=7cm,height=5cm]{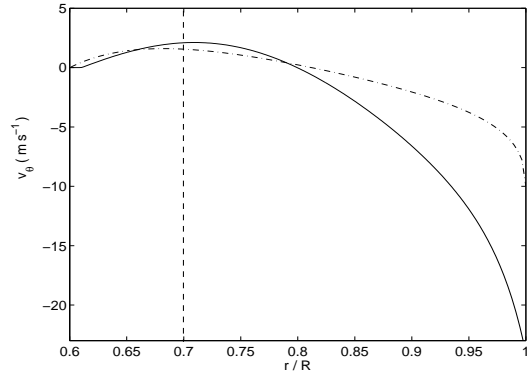}}
\caption{Plot of $v_{\theta}$ (in m s$^{-1}$) as a function of $r$ at the
mid-latitude $\theta = 45^{\circ}$. The solid line shows the velocity
used by CNC, whereas the dashed dotted line is the velocity used by Dikpati
\& Charbonneau (1999; see Fig.~1c).}   
\end{figure}

Defending the meridional circulation used by
Dikpati \& Charbonneau (1999), DRGM write: ``Given the knowledge
available at that time about the meridional circulation, it was
reasonable to try what Dikpati \& Charbonneau (1999) did, but not
in 2004 \ldots''
DRGM have not clarified for the benefit of the readers
whether it was still reasonable to
try this in 2002 when Nandy \& Choudhuri (2002) wrote their paper,
although during 2001--2002 it was stated in successive dynamo
papers of the HAO group (Dikpati \& Gilman 2001; Dikpati et al.\
2002) that they were using the same meridional circulation as Dikpati \&
Charbonneau (1999)!
However, we are given to understand that 2004 is the decisive year
when such meridional circulations
ceased to be `reasonable'---the year marked by the publication
of a paper by Gilman \& Miesch (2004). 
We humbly beg to differ from this point of view. The bottom of the
SCZ is the least understood region in the interior of the Sun.  Recently
one of us (A.R.C.) attended a workshop on the tachocline at Isaac Newton
Institute in Cambridge, where one of the authors of the DRGM paper (P.A.G.)
was also present.  From the very heated discussions there, it was obvious
that there is no general agreement regarding the extent of overshooting
below the base of SCZ or the extent of turbulence in the tachocline. We
agree with Gilman \& Miesch (2004) that the meridional circulation could
not penetrate much into a stable region where there is no overshooting
or turbulence. However, in our current state of ignorance, we cannot
rule out the possibility of enough overshooting and turbulence
existing throughout the tachocline below the bottom of SCZ and a
meridional circulation penetrating through this region. We cannot
present detailed arguments within the three pages kindly allotted by
the A\&A Editor for our reply. A paper under preparation will address
this issue.  

\section{Conclusion}

To sum up, DRGM probably got results different from ours 
by treating the magnetic
buoyancy unsatisfactorily.  Their other criticisms of our model do
not appear very relevant.

\section*{References}
Chatterjee, P., Nandy, D., \& Choudhuri, A.\ R. 2004, A \& A, 427, 1019 \\
Choudhuri, A.\ R. 1998, The Physics of Fluids and Plasmas: An Introduction
for Astrophysicists (Cambridge: Cambridge University Press)\\
Choudhuri, A.\ R. 2003 Sol.\ Phys., 215, 31\\
Choudhuri, A.\ R., Chatterjee, P., \& Nandy, D. 2004, ApJ, 615, L57 \\ 
Choudhuri, A.\ R., \& Dikpati, M.  1999, Sol.\ Phys., 184, 61 \\
Choudhuri, A.\ R., Sch\"ussler, M., \& Dikpati, M. 1995 A \& A, 303, L29\\
Dikpati, M., \& Charbonneau, P. 1999, ApJ, 518, 508 \\
Dikpati, M., \& Choudhuri, A.\ R. 1995 Sol.\ Phys., 161, 9\\
Dikpati, M., Corbard, T., Thompson, M.\ J., \& Gilman, P.\ A. 2002, ApJ, 575, L41 \\
Dikpati, M., \& Gilman, P.\ A. 2001, ApJ, 559, 428\\ 
Dikpati, M., Rempel, M., Gilman, P.\ A., \& MacGregor, K.\ B. 2005, A \& A, in press \\
Gilman, P.\ A., \& Miesch, M.\ S. 2004, ApJ, 611, 568 \\
Nandy, D., \& Choudhuri, A.\ R. 2001, ApJ, 551, 576\\
Nandy, D., \& Choudhuri, A.\ R. 2002, Science, 296, 1671 \\
\end{document}